\documentclass[preprint,showpacs,pra,aps,amsmath,amssymb,unsortedaddress]{revtex4}
\usepackage[dvips]{graphicx}
\usepackage{dcolumn}
\usepackage{amsmath}
\usepackage{times}

\begin{document}

\draft

\title{On the Coulomb-Sturmian matrix elements of the Coulomb Green's operator}
\author{F.\ Demir, Z.\ T.\ Hlousek and Z.\ Papp}
\affiliation{ Department of Physics and Astronomy, California State
University, Long Beach, California 90840 }

\date{\today}

\begin{abstract}\noindent
The two-body Coulomb Hamiltonian, when calculated in Coulomb-Sturmian basis, has an infinite symmetric tridiagonal form, also known as Jacobi matrix form. This Jacobi matrix structure involves a continued fraction representation for the inverse of the Green's matrix. The continued fraction can be transformed to a ratio of two $_{2}F_{1}$ hypergeometric functions. From this result we find an exact analytic formula for the matrix elements of the Green's operator of the Coulomb Hamiltonian. 
\end{abstract}

\maketitle

\section{Introduction}

Dynamics of a quantum system is fully determined by its Green's operator. The knowledge of Green's operator is equivalent to the complete understanding of the system. From the Green's operator we can extract the complete spectrum and the wave functions.

However, in most of the cases, it is not possible to calculate the exact Green's operartor. Fortunately, it is often sufficient to determine the Green's operator of an asymptotic Hamiltonian, because the rest can be considered as a perturbation and can be approximated by finite matrices. This way, having an analytic representation of the Green's operator of some asymptotic Hamiltonian, one can build a powerful quantum mechanical approximation method.

This is the central idea behind an approximation scheme that is used to solve few-body problems. 
In particular, we have applied this scheme with great success to determine the solution three-body Faddeev equations  with Coulomb interactions (see eg. \cite{faddeev-papers} and references therein). 
The success of the calculation depended very much on our ability to evaluate the matrix elements of the Coulomb Green's operator between Coulomb-Sturmian basis states. The choice of Coulomb-Sturmian basis is essential. In that basis, the two-body Coulomb Hamiltonian has  Jacobi-matrix  ($J$-matrix) form. As a result, the matrix elements of the Green's operator satisfy three-term recursion relations that greatly simplify the calculation of the matrix elements of the Green's operator of the Coulomb Hamiltonian. In Ref.\ \cite{klp} we have described two independent ways to calculate the matrix elements of the Coulomb Green's operator. They are suitable for different regions of the complex energy plane. In the first method, the J-matrix was used as a three-term recursion relation. The seed, which is related to $_{2}F_{1}$ hypergeometric function, was derived from independent considerations. In the second method, the explicit inversion of the infinite J-matrix resulted in a continued fraction. We also note another method \cite{heller} for calculating Green's matrices which is based on the solution of scattering problems on $L^{2}$ basis \cite{heller-jamani}.

In this article we unify the two approaches presented in Ref.\ \cite{klp}. 
We start from the continued fraction representation for the ratio of two consecutive matrix elements of Coulomb Green's operator. 
A closer inspection reveals that this continued fraction corresponds to the continued fraction of the ratio of two $_{2}F_{1}$ hypergeometric functions. From this we obtain an exact analytic formula for the Coulomb Greens matrix. 

In Section II we derive an expression for the inverse of the $N\times N$ Green's matrix. A factor $C_{N+1}$ in the inverse of the truncated Green's matrix is expressed as a continued fraction. 
In section III, we apply our formulae to a Coulomb problem.  We show that $C_N$ is exactly computable. We observe that the continued fraction $C_N$ is identical to a certain $T$-fraction that corresponds to a ratio of two hypergeometric functions $_{2}F_{1}$. Establishing the equivalence amounts to solving a system of five equations and choosing proper region of convergence of the representation. Finally, we establish formulae that are used to compute all elements of the Green's matrix.

\section{Continued fraction representation of the Green's operator}

Formally, the Green's operator $G$ is defined by the equation,
\begin{equation}\label{eq:gf_def}
G(z)\left( z- H\right) = \left(z-H\right) G(z) = 1~,
\end{equation}
where $z$ is a complex number and $H$ is the Hamiltonian.

Suppose that the operator $z-H$, evaluated in some discrete Hilbert-space basis
$\{| i \rangle \}$, has an infinite symmetric tri-diagonal, i.e. Jacobi-matrix, structure,
\begin{equation}\label{eq:tri-diag-H}
  z-H \equiv J = \left(\begin{matrix}
J_{0,0} & J_{0,1} & 0       & 0       & \ldots \\
J_{1,0} & J_{1,1} & J_{1,2} & 0       & \ldots \\
0       & J_{2,1} & J_{2,2} & J_{2,3} & \ldots  \\
\vdots & \ddots & \ddots & \ddots & \ddots \\
\end{matrix}\right)~.
\end{equation}
In such a basis, Eq.\  (\ref{eq:gf_def}) becomes
\begin{equation}\label{eq:tri-diag-HG}
\left(\begin{matrix}
J_{0,0} & J_{0,1} & 0       & 0       & \ldots \\
J_{1,0} & J_{1,1} & J_{1 2} & 0       & \ldots \\
0       & J_{2,1} & J_{2,2} & J_{2,3} & \ldots  \\
0       & 0       & J_{3,2} & J_{3,3} & \ldots  \\
\vdots & \ddots & \ddots & \ddots & \ddots  \\
\end{matrix}\right)
\left(\begin{matrix}
G_{0,0} & G_{0,1} & G_{0,2} & G_{0,3} & \ldots \\
G_{1,0} & G_{1,1} & G_{1,2} & G_{1,3} & \ldots \\
G_{2,0} & G_{2,1} & G_{2,2} & G_{2,3} & \ldots  \\
G_{3,0} & G_{3,1} & G_{3,2} & G_{3,3} & \ldots  \\
\vdots & \ddots & \ddots & \ddots & \ddots  \\
\end{matrix}\right)
=
\left(\begin{matrix}
1 & 0 & 0  & 0 & \ldots \\
0 & 1 & 0  & 0 & \ldots \\
0 & 0 & 1 & 0 & \ldots  \\
0 & 0 & 0 & 1 & \ldots  \\
\vdots & \ddots & \ddots & \ddots & \ddots  \\
\end{matrix}\right)~.
\end{equation}

The knowledge of the $N\times N$ upper left corner of the full Green's matrix is sufficient to determine the physical quantities. Let us denote the corresponding $N\times N$ upper left corner matrices by $J^{(N)}$, $G^{(N)}$ and $1^{(N)}$, respectively.
If we multiply the $N\times \infty$ part of $J$ with the $\infty \times N$
part of $G$ we get the $N\times N$ unit matrix $1^{(N)}$. 
The sum, due to the tridiagonality of $J$, is reduced to three terms
\begin{equation}\label{eq:gf-eq-2solve}
J_{n,n-1}G_{n-1,m} + J_{n,n}G_{n,m} + J_{n,n+1}G_{n+1,m} = \delta_{n,m}~,
\end{equation}
where $n=0,1,..N$ and $m=0,1,..N$. If $n<N$, 
only terms from $G^{(N)}$ are appearing in the sum. For the $n=N$ case, we have:
\begin{equation}\label{eq:gf-eq-2solveN}
J_{N,N-1} G_{N-1,m} + J_{N,N}G_{N,m} + J_{N,N+1} G_{N+1,m} = \delta_{N,m}~.
\end{equation}
The $G_{N+1,m}$ elements are outside the $G^{(N)}$ matrix. 
We can eliminate them formally by writing
\begin{equation}\label{eq:gf-eq-2solveN2}
J_{N,N-1} G_{N-1,m} + ( J_{N,N} + J_{N,N+1} G_{N+1,m}/ G_{N,m} ) G_{N,m} = \delta_{N,m}~.
\end{equation}
This formal elimination of the elements outside of $G^{(N)}$ amounts to modifying single element, $J_{N,N}$ of $J^{(N)}$.

We can calculate the ratio $G_{N+1,m}/ G_{N,m}$ from another, independent
relation:
\begin{equation}\label{eq:gf-eq-2solveN}
J_{N+1,N} G_{N,m} + J_{N+1,N+1}G_{N+1,m} + J_{N+1,N+2} G_{N+2,m} = 0~.
\end{equation}
By rearranging, we get
\begin{equation}\label{eq:cf001}
\left(-\cfrac{1}{J_{N+1,N}}\cfrac{G_{N+1,m}}{G_{N,m}} \right) = \cfrac{1}{ J_{N+1,N+1} - J_{N+1,N+2}
\left( -\cfrac{1}{J_{N+2,N+1}}\cfrac{G_{N+2,m}}{G_{N+1,m}}\right)J_{N+2,N+1} }~.
\end{equation}
We introduce a simplifying notation
\begin{equation}\label{eq:cf01}
C_{N+1} = -\cfrac{1}{J_{N+1,N}}\cfrac{G_{N+1,m}}{G_{N,m}}~.
\end{equation}
Then, equation (\ref{eq:cf001}) becomes:
\begin{equation}\label{cn}
C_{N+1} = \cfrac{1}{ J_{N+1,N+1} - J_{N+1,N+2}  C_{N+2}   J_{N+2,N+1} }~,
\end{equation}
or
\begin{equation}\label{cn-1}
C_{N+1}^{-1} = { J_{N+1,N+1} - J_{N+1,N+2}  \cfrac{1}{C_{N+2}^{-1}}   J_{N+2,N+1} }~.
\end{equation}
 A repeated application of this relation 
results in a continued fraction. Taking into account that the J-matrix is symmetric,
we get
\begin{equation}\label{eq:cf00}
C_{N+1}^{-1} =  
{J_{N+1,N+1} - \cfrac{ J_{N+1,N+2}^2 }{J_{N+2,N+2} - 
 \cfrac{  J_{N+2,N+3}^2 } { J_{N+3,N+3} - 
 \cfrac{J_{N+3,N+4}^2 } {\ddots}} }}~.
\end{equation}
 This continued fraction does not depend on the index $m$, the correction term to $J_{N,N}$ is the same for all $m$'s.  Consequently, we can write
Eq.\ (\ref{eq:gf-eq-2solveN2}) in the form
\begin{equation}
( {J}^{(N)}_{i,j} - \delta_{i,N} \delta_{j,N} J_{N,N+1}^{2} C_{N+1} ) G^{(N)}=1^{(N)}.
\end{equation}
The modified Jacobi matrix, $ {J}^{(N)}_{i,j} - \delta_{i,N} \delta_{j,N} J^2_{N,N+1} C_{N+1}$, is  
the inverse matrix of $G^{(N)}$
\begin{equation}
G^{(N)}=  ( {J}^{(N)}_{i,j} - \delta_{i,N} \delta_{j,N} J_{N,N+1}^{2}  C_{N+1} )^{-1} ~.
\end{equation}

\section{$D$-dimensional Coulomb problem}

The Hamiltonian of the $D$-dimensional Coulomb problem, with $ D\geq 2$, is given by 
\begin{equation}\label{eq:coulomb-h}
H = - \frac{\hbar^2}{2m} \left( \frac{d^2}{d r^2} - \frac{  L(L+1)  }{r^{2}} \right)   + \frac{Z}{r}~,
\end{equation}
where  $L = l +\frac{D-3}{2}$. In the calculation below we set $\hbar=m=1$.
The Coulomb-Sturmian functions are defined by
\begin{equation}\label{eq:CS-basis}
\psi_n(r) =  \left[ \frac{\Gamma(n+1)}{\Gamma(n+2L+2)}\right]^{1/2}
\hbox{\rm e}^{-b_S r} (2 b_S r)^{L+1} L_n^{2L+1}(2 b_S r)~,
\end{equation}
where $n=0,1,\cdots$,  ${L}^\alpha_{n}$ is an associated Laguerre polynomial and $b_S$ is a parameter. 
The Coulomb-Sturmian functions form a discrete basis, on which the Coulomb Hamiltonian has a  $J$-matrix form \cite{klp}
\begin{equation}
J_{n,m}=\left\{
\begin{matrix}
\frac{ \displaystyle  k^2-b_S^2}{ \displaystyle  2b_S}(n+L+1) - Z & \textrm{for}~~~~ n=m~, \\
 -\frac{ \displaystyle  k^2+b_S^2}{\displaystyle   4b_S}\sqrt{(n+1)(n+2L+2)} & \textrm{for}~~~~ m=n+1~, \\
 -\frac{  \displaystyle k^2+b_S^2}{ \displaystyle  4b_S} \sqrt{n(n+2L+1)} & \textrm{for}~~~~ m=n-1~, \\
\end{matrix}
\right.
\end{equation}
where $k=\sqrt{2z}$. 

To calculate the Green's matrix of Coulomb Hamiltonian we need to compute the continued fraction $C_N$.
According to (\ref{cn-1}) and (\ref{eq:cf00}),  the continued fraction $C_N$ satisfies the relation
\begin{equation}\label{cfrac}
\begin{split}
C_N^{-1} = & \left[ \frac{ \displaystyle  k^2-b_S^2}{  \displaystyle 2b_S} (N+L+1) - Z\right] +
(-)\frac{ \displaystyle  k^2+b_S^2}{ \displaystyle  4b_S} (N+1)(N+2L+2) C_{N+1} \\
 = & \left[ \frac{  \displaystyle  k^2-b_S^2}{  \displaystyle 2b_S} (N+L+1) - Z  \right]  +
\mathop{\mathbf{K}}\limits_{p=1}^\infty
\left(
\frac{-\left(\frac{\displaystyle   k^2+b_S^2}{ \displaystyle  4b_S}\right)^2 (N+p)(N+2L+1+p)} 
{ \frac{ \displaystyle  k^2-b_S^2}{ \displaystyle  2b_S}  (N+L+1+p)  -Z }
\right)~.    \\
\end{split}
\end{equation}

From the theory of continued fractions  we know that a particular ratio of two hypergeometric functions can be 
represented as a continued fraction from the class of so called $T$-fractions \cite{l-w}
\begin{equation}\label{eq:tg-fraction}
T(a,b;c;y) =
\left( c+\left(b-a+1\right)y\right) + \mathop{\mathbf{K}}\limits_{p=1}^\infty
\left( \frac{-\left( c-a+p\right)\left(b+p\right) y}{c+p+\left(b-a+1+p\right)y}\right) =
c\frac{ {_2}F{_1} (a,b;c;y)}{ {_2}F{_1}(a,b+1;c+1;y)}~.
\end{equation}
This $T$-fraction converges to the ratio of two hypergeometric functions if $\vert y\vert <1$. Also, for $y=-1$, the representation is convergent if $\vert \Im (c-a+b)\vert < \vert \Re (c+a-b-1)\vert$.

By comparison, we see that two continued fractions, Eq. (\ref{cfrac})  and Eq. (\ref{eq:tg-fraction}) have identical structure. 
Both fractions have numerators that are quadratic in index $p$ and denominators that are linear in index $p$. Lorentzen and Waadeland \cite{l-w} have shown that all $T$-fractions of this type are convergent and expressible as a ratio of two hypergeometric functions.

Hence, we can write,
\begin{equation}
T(a,b;c;y) = c\frac{{_2}F{_1}(a,b;c;y)}{{_2}F{_1}(a,b+1;c+1;y)} = d C_N^{-1}~,
\end{equation}
where $d$ is an overall scale parameter.
Parameters, $a,b,c,y$ and $d$ are determined from the set of five equations subject to a
convergence condition $\vert y\vert <1$.
The five equations that determine parameters are:
\begin{eqnarray}
 y = & d^2 \left( \frac{\displaystyle{k^2+b_S^2}}{\displaystyle{4b_S}} \right)^2 ~,\\
 1+y = &  d \left( \frac{\displaystyle{k^2-b_S^2}}{\displaystyle{2b_S}}\right) ~, \\
 y(b+c-a) = & d^2 \left(  \frac{\displaystyle{k^2+b_S^2}}{\displaystyle{4b_S}}\right)^2 (2 N+2 L+1) ~, \\
 yb(c-a) = &  d^2 \left(  \frac{\displaystyle{k^2+b_S^2}}{\displaystyle{4b_S}}\right)^2 N(N+2L+1) ~, \\
 c+ (b-a +1) y = & d \left( -Z + \left(\frac{\displaystyle{k^2-b_S^2}}{\displaystyle{2b_S}}\right) (N+L+1) \right) ~.
\end{eqnarray}

From the first two equations, together with the convergence condition $\vert y\vert <  1$, we get
\begin{eqnarray} 
d = & - \frac{ \displaystyle 4 b_{S}}{\displaystyle \left( b_{S} -ik\right)^2
} ~,\\
y = & \left(\displaystyle \frac{ {\displaystyle b_{S} } +ik }{ \displaystyle b_{S} -ik }\right)^2~.
\end{eqnarray}
There are two sets of solutions for the remaining parameters, (we also introduce the conventional parameter $\gamma = Z/k$)
\begin{eqnarray}
\left\{
\begin{matrix}
 a = - L +   \mathrm{i}  \gamma \\  b = N \\  c = N+L+1 +  \mathrm{i}   \gamma \\
  \end{matrix}
\right.
~~~~~~~
\left\{\begin{matrix}  a =  L + 1 +  \mathrm{i}  \gamma \\  b = N +2L +1\\  c = N+L+1 +  \mathrm{i}  \gamma \end{matrix}  \right.
\end{eqnarray}
In fact, the two solutions yield the same expression for $C_N$. This is easy to show using the identity
satisfied by the hypergeometric function:
\begin{equation}
{_2}F_{1}(a , b; c ;y) = {_2}F{_1}(c-b,c -a ; c;y) \times
{_2}F{_1}(a, a + b - c; a;y)
\end{equation}
In our case this identity reads,
\begin{equation}
{_2}F{_1}(-L+  \mathrm{i}  \gamma,N;N+L+1+  \mathrm{i}   \gamma;y) =
\frac{\displaystyle{ {_2}F{_1}(L+1+   \mathrm{i}  \gamma,N+2L+1;N+L+1+   \mathrm{i}  \gamma;y)}} 
{\displaystyle{\left( 1-y\right)^{(2L+1)}}}~.
\end{equation}
Finally, we have the following closed form expression for the continued fraction 
\begin{equation}\label{eq:coulomb-cf-final}
C_{N} =
\frac{ - \frac{ \displaystyle  4 b_{S}} 
{ \displaystyle    \left({\displaystyle  b_{S} -   \mathrm{i}  k  }\right)^{2}} }
{ \displaystyle  N+L+1 +   \mathrm{i}  \gamma } \;
\frac{{_2}F{_1}\left( \displaystyle -L+   \mathrm{i}  \gamma  , N+1; N+L+2 +   \mathrm{i}  \gamma   ;
\left(\frac{{b_{S}}  + \mathrm{i} k  }   {{b_{S}} -   \mathrm{i}  k  }\right)^2\right)}
{{_2}F{_1}\left( \displaystyle -L+  \mathrm{i}  \gamma , N; N+L+1 +  \mathrm{i}  \gamma ;
\left(\frac{ { b_{S}} + \mathrm{i} k }{ {  b_{S}} -   \mathrm{i} k }\right)^2\right)}~.
\end{equation}

The complete discrete spectrum of the system follows from any matrix element of the Green's operator and they are constructed by recursion relations, see equations (\ref{eq:cf01}) and  (\ref{cn-1}). In particular,
\begin{equation}
\begin{split}
\displaystyle G_{N,m} = & -C_{N}J_{N,N-1}G_{N-1,m}\\
C_{N+1} = & \frac{J_{N,N}}{J^2_{N,N+1}} - \frac{1}{J^2_{N,N+1}} C^{-1}_{N}\\
\end{split}
\end{equation}
For example, $G_{0,0} = C_0$,
\begin{equation}\label{eq:G00-coulomb}
G_{0,0} = 
-\frac{4 b_{S}}{(b_{S}-\mathrm{i}k)^{2}} \:
\frac{1}{L+1 + \mathrm{i} \gamma } \;
{_2}F{_1}\left(-L + \mathrm{i} \gamma , 1; L+ 2+  \mathrm{i} \gamma  ;
\left( \frac{b_{S} + \mathrm{i} k}{b_{S} - \mathrm{i} k}\right)^{2}  \right)~.
\end{equation}

The ${_2}F{_1}$ hypergeometric function has branch-cut singularity on the $\left[1,\infty \right)$ interval.  This cut is mapped in Eqs.\  (\ref{eq:coulomb-cf-final}) and (\ref{eq:G00-coulomb})  to a branch cut in the $z$-plane along the positive real axis. In the case of the attractive Coulomb potential $Z<0$, and $\mathrm{i}\gamma = \mathrm{i}Z/k$ can take negative values. Then, there is a first order pole on the negative real axis at $L+1 + \mathrm{i}\gamma =0$.  Additional poles are determined from the hypergeometric function by the condition,
$ L+2 + \mathrm{i}\gamma = 0,-1,-2,\ldots$. All together,  there is a set of infinitely many discrete first order poles given  by the formula,
\begin{equation}
z_{n_r} = E  = \frac{1}{2} \left( \frac{Z}{ \mathrm{i}\gamma}\right)^2= -\frac{1}{2} \frac{Z^2 }{(n_r+L+1)^2}~,~~~~ n_r = 0,1,2,\ldots ~.
\end{equation}
This is a complete discrete spectrum of the attractive Coulomb system.

\section{Summary}

In this paper we derived a closed formula for the Coulomb-Sturmian matrix elements of the Coulomb Green's operator. In particular, we expressed the inverse of the $N\times N$ Coulomb Green's matrix in terms of ${_2}F{_1}$ hypergeometric functions. The ratio of these two 
${_2}F{_1}$ hypergeometric function can easily be evaluated in numerical computations by another Gauss-type continued fraction \cite{f21calc}.



\begin{thebibliography}{99}

\bibitem{faddeev-papers}  Z.\ Papp,  C-.Y.\ Hu, Z.\ T.\ Hlousek, B. K\'onya and S.\ L.\ Yakovlev,
Phys.\ Rev.\ A,  {\bf 63}, 062721 (2001).

\bibitem{klp} B.\ K\'onya, G.\ L\'evai, and Z.\ Papp, 
J.\ Math.\ Phys. {\bf 38}, 4832 (1997).


\bibitem{heller}  E.~J.~Heller, Phys. Rev. A {\bf 12}, 1222 (1975).

\bibitem{heller-jamani}  E.~J.~Heller and H.~A.~Jamani, Phys. Rev. A {\bf 9}, 1201 (1974); ibid. {\bf 9}, 1209 (1974); H.~A.~Jamani and L.~Fishman, J. Math. Phys {\bf 16}, 410 (1975)


\bibitem{l-w} L. Lorentzen and H. Waadeland, Continued Fractions with Applications.
Studies in Computational Mathematics 3. North-Holland (1992). page 308.

\bibitem{f21calc} Ref.\ \cite{l-w}, page 293-301.

\end{thebibliography}
\end{document}